\documentclass[trackchanges, twocolumn]{aastex701}
\usepackage{float}
\usepackage{url}
\usepackage{ulem}

\begin{document}

\title{A New Approach for Testing Einstein's Theory of Gravity Close to Rapidly Spinning Black Holes}

\author[orcid=0009-0005-0818-7484,sname='Menon', gname=Shravan Vengalil]{Shravan Vengalil Menon}
\affiliation{Department of Physics, Washington University in St. Louis, 1 Brookings Dr, St. Louis, MO 63130, USA.}
\email[show]{s.vengalilmenon@wustl.edu} 

\author[orcid=0000-0002-9705-7948,sname='Hu',gname=Kun]{Kun Hu}
\affiliation{Department of Physics, Washington University in St. Louis, 1 Brookings Dr, St. Louis, MO 63130, USA.}
\email[show]{hkun@wustl.edu} 

\author[0000-0002-1084-6507,sname=Krawczynski,gname=Henric]{Henric Krawczynski}
\affiliation{Department of Physics, Washington University in St. Louis, 1 Brookings Dr, St. Louis, MO 63130, USA.}
\email[show]{\\krawcz@wustl.edu}

\begin{abstract}

The Penrose process and the collisional Penrose process involve particles decaying or interacting very close to a spinning black hole, respectively, during which some particles are pushed into negative energy trajectories and fall into the black hole while others gain that energy and escape the system.
These two processes are difficult to observe as they occur very rarely 
by chance. Here we report a new observational signature of similar, but less extreme processes occurring in and near the ergospheres of rapidly spinning black holes. We find that the reflection of the thermal emission from a  geometrically thin, optically thick accretion disk can lead to the formation of a power-law component, even in the absence of a corona. Unlike the well-known Penrose processes, the scattering particles lose a good fraction of their energy, but are not pushed into negative energy trajectories. 
We emphasize that this new component has distinct spectral and 
polarimetric properties that can be used for its identification 
as long as it out-competes other power-law emission components.
We emphasize that the new component needs to be taken into account 
when interpreting spectral and spectropolarimetric observations of black holes.
The detection and unambiguous identification of the new component with current
or future  broadband X-ray spectral and spectropolarimetric missions 
can open a new window into testing Einstein's theory of gravity close 
to the edges of black holes, and opens up new opportunities to 
constrain black hole spins and inclinations.
\end{abstract}

\keywords{\uat{Black holes}{162} --- \uat{General Relativity}{641} --- \uat{High Energy astrophysics}{739} --- \uat{X-ray astronomy}{1810}}


\section{Introduction} 
A Black Hole X-ray Binary (BHXRB) comprises a black hole that accretes matter from its binary star companion. 
The matter falling towards the black hole forms an accretion disk that transports mass towards the black hole while shedding energy and angular momentum via radiation and focused as well as unfocused outflows \citep{2019arXiv191104305F}.
BHXRBs are observed 
in distinct spectral states which reflect different configurations of the accreting matter. In the soft state, the emission is dominated by thermal emission from a geometrically thin optically thick accretion disk extending close to the innermost stable circular orbit (ISCO) of the black hole \citep{ShakuraSunyaev}. 
A disk at a temperature of about 10 million K close to the ISCO emits X-rays with keV (kilo-electron volt) energies. As the disk exhibits a temperature profile peaking close to the black hole, the emitted energy spectrum exhibits a multi-temperature blackbody energy spectrum. On the other hand, the hard state is characterized by repeated inverse Compton scatterings of  X-rays producing a power-law component in the observed spectra. Inverse Compton scattering refers to photons gaining energy when colliding with relativistically moving particles or plasma. The hot plasma energizing the photons is commonly referred to as ``the corona'' \citep[e.g., see][and references therein]{2025ApJ...993...54K}. The Comptonization of the emission, i.e.\ the energization through repeated inverse Compton scatterings, may be effected by electrons or positrons in very hot plasma \citep{Shapiro76,Katz76,corona}, or by bulk plasma moving at mildly relativistic velocities \citep{Beloborodov2017,SironiBeloborodov2020,Sridhar2025,2025ApJ...993...54K}.
Some of the coronal emission reflects off the highly ionized accretion disk giving rise to an emission component called the reflected emission \citep{reflection,2026arXiv260103349N}. 
Independent of the emission state, the matter close to the black hole is highly ionized and highly reflecting \citep{Done_2007,2022ApJ...934....4K,2026arXiv260103349N}.

In this paper, we report the results of model prediction of a new power-law emission component
in the soft state of rapidly spinning black holes that dominates the thermal emission at energies well above the thermal cutoff. The new emission component exhibits a  power-law energy spectrum ($dN_{\gamma}/dE\propto E^{-\Gamma}$ with $N$ the number of emitted photons and $E$ the energy of the photons). 
The index $\Gamma$ depends on the  black hole spin parameter 
and on the observer inclination. 
The component arises from the reflection of the gravitationally lensed thermally emitted accretion disk photons returning to the accretion disk 
and experiencing large energy gains in one or several key scattering processes. The new power-law component forms even in the complete absence of any coronal plasma.
The new component was not seen in previous theoretical studies of the emission from geometrically thin, optically thick black hole accretion disks as it requires high black hole spins and is missed by studies that do not account 
for the kinematics of the photon scatterings in the Kerr space time
\citep[see e.g.,][]{2005ApJS..157..335L}.
In the rest of this paper, we discuss the methods used to obtain the result (Section \ref{sec:2}), the nature of the effect (Section \ref{sec:3}) and its broader implications (Section \ref{sec:4}).


\section{Methods} \label{sec:2}

In the following, we describe the {\tt kerrC} ray tracing code used in our simulations. The {\tt kerrC} code is based on the {\tt xTrack} code \citep{2012ApJ...754..133K}. We model the propagation of photon packages in Kerr spacetime characterized by the metric $g_{\mu\nu}$ in Boyer-Lindquist (BL) coordinates $x^{\mu} = (t, r, \theta, \phi)$. We wrote here ``photon packages'', 
to emphasize that we use Stokes parameters to characterize the polarization information carried by a statistical ensemble of photons.
In the following, we simply refer to these photon packages as photons. 
All photons originate as thermal emission from a multi-temperature accretion disk \citep{1974ApJ...191..499P}. 
The photon energies are drawn from 
a blackbody distribution at a temperature 
that is a factor of 1.8 higher than 
the local disk temperature. 
This hardening of the emission 
parametrizes the effect of the interaction 
of the photons in the
accretion disk atmosphere \citep{1995ApJ...445..780S,2019ApJ...874...23D}.
The photon trajectories are calculated by numerically integrating the geodesic equation with a 4.5th-order Cash-Karp integrator. The linear polarization four vectors are parallel transported along the geodesics.
We fix the black hole spin to $a = 0.998$ and adopt a temperature scaling factor $\sigma \equiv (L/10\%L_{Edd})^{1/4}(M/10M_{\odot})^{-1/4}$ to 0.75. For Cyg X-1 black hole mass $M=21.2 M_\odot$, this choice of $\sigma$ corresponds to $L\sim6.7\%L_{Edd}$.

Photons are emitted and scattered in the reference frame comoving with the plasma. We utilize orthonormal tetrads $e^{\mu}_{(\nu)}$ to convert the photon wave vectors and polarization vectors  between the different frames. 
%
The initial polarization of each photon is assigned with the tabulated values of Chandrasekhar for the emission of a pure indefinitely deep scattering atmosphere and  the scattering uses Chandrasekhar's formalism for the reflection of polarized light off an indefinitely deep  scattering atmosphere \citep{chandrasekhar}.
The formalism assumes a 100\% disk reflectivity and scattering in the Thomson regime, neglecting the energy exchange between the photon and electron. 
The approximation of 100\% reflectivity should hold 
in good approximation in the very inner region of the accretion flow of highly spinning black holes.
Our treatment neglects the temperature of the reflecting electrons as well as the heating and cooling of the accretion disk from the scattering of the returning radiation. We will return to this latter topic 
in Section \ref{sec:4}.
The scattering is simulated as follows.
First we draw a random direction of the scattered beam. The polarization degree and vector of the incoming photon in the plasma rest frame are subsequently translated into a Stokes $I$ parameter (intensity), and Stokes $Q$ and $U$ parameters (polarization degree and angle) referenced to the scattering plane. Subsequently, Chandrasekhar's scattering matrices are used to calculate the Stokes parameters of the outgoing photons, accounting for all scattering orders. 
The Stokes parameters are then converted back into the polarization vector and polarization degree of the outgoing photon. Note that the  change of Stokes $I$ averages to 1 over all scatterings that the code performs corresponding 
to a 100\% disk scattering efficiency 
\citep{2012ApJ...754..133K}.
Once the photon reaches a distance of $R = 10,000 \, r_g$, the wave vector is transformed into the coordinate stationary frame of the observer, and the energy at infinity and the polarization of the photon are extracted. Detailed descriptions of the inner workings of the code can be found in \cite{2012ApJ...754..133K}.



\section{Results}
\label{sec:3}
In the following, we show results obtained for a nearly maximally rotating black hole with the dimensionless spin parameter fixed at $a=0.998$ (Here, $a\,=\,cJ/GM^2$ with $c$ the speed of light, $J$ the angular momentum of the black hole, $G$ Newton's gravitational constant, and $M$ the mass of the black hole). We assume that the accretion disk extends all the way to the ISCO. We show the results for an observer inclination of 33.6$^{\circ}$ close to the most likely value of the bright and well studied BHXRB Cygnus X-1 \citep{2025ApJ...993...54K}.
\begin{figure}[ht!]
\centering
\includegraphics[width=0.47\textwidth]{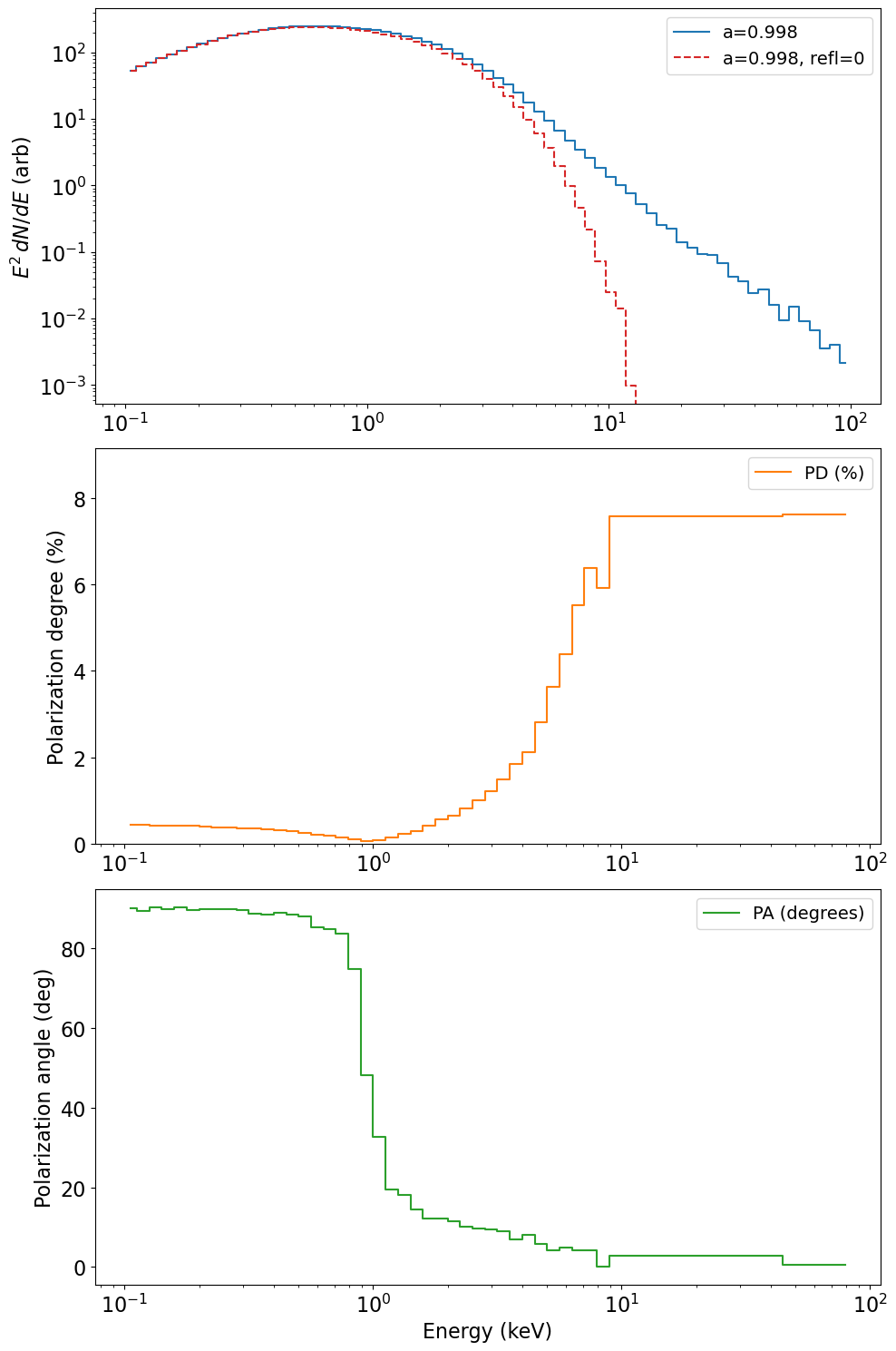}
\caption{Spectral energy distribution (SED, $E^2 dN/dE$ with $N$ and $E$ being the number and energy of the photons, respectively) for a simulated black hole in the soft state with a spin parameter $a = 0.998$, along with its polarization degree (PD, middle panel) and angle (PA, bottom panel) for the total signal. The blue line shows the results for a 
geometrically thin, optically thick, highly ionized (100\% reflectivity) accretion disk seen at an inclination of 33.6$^{\circ}$. 
The red dotted curve shows the results when the reflection is switched off 
in the simulations. Neither model has a corona.}
\label{spectrum}
\end{figure}

We find that the scattering of the gravitationally lensed thermally emitted disk photons can produce a power-law component 
that strongly dominates over the multi-temperature blackbody emission 
at and above the Wien tail of the thermal disk emission. 
For the specific simulations that we show, 
this power-law component starts to be noticeable at $\sim$2\,keV, 
becomes increasingly more important at higher energies, 
and dominates the emission at $10$\,keV and above 
(Figure~\ref{spectrum}, top panel).
Our simulations show that the power-law slope and cutoff depend on the black hole spin. 
Our modeling makes some simplifications, i.e., it neglects the effect
of scatterings in the Compton (or even Klein-Nishina) rather than 
in the Thompson regime. Furthermore, we do not account for the 
impact of the returning emission on the radial temperature 
profile of the accretion disk.

\begin{figure*}
\centering
\includegraphics[width=0.8\textwidth]{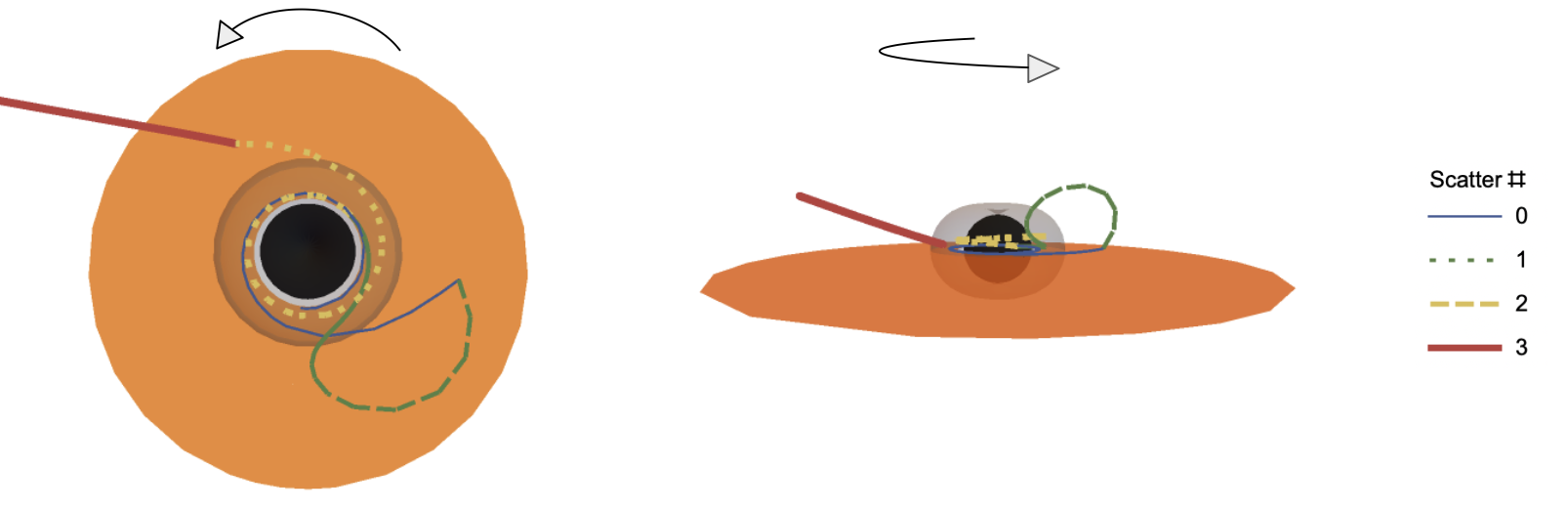}
\caption{Trajectory of a photon that comprises a power-law event along with the black hole and the accretion disk (orange) when seeing the disk face on (left panel) and edge on (right panel). The line colors show the energy at infinity of the photon after being launched ($1.7$ keV, thin blue line), scattering once ($0.7$ keV, dotted green line), twice ($28.6$ keV, dashed yellow line), and three times ($28.8$ keV, thick red line). The first two scatterings occur within the ergosphere shown in gray.}

\label{intensity}
\end{figure*}

Figure \ref{intensity} shows for illustrative purposes the trajectory of one of the photons deep in the tail of the power-law emission component.  
The following discussion uses the photon energy at infinity 
\begin{equation}
E_\infty = -p_t =-g_{t\alpha} p^{\alpha}   
\label{e:i}
\end{equation}
with $p_t$ being the covariant time component of the photon's 
four-momentum $p^\alpha$, and $g_{0\alpha}$ being the contravariant 
time components of the metric. The energy at infinity is conserved 
along geodesics. For photons escaping the black hole's gravitational attraction, it equals the photon energy measured by an observer 
in the asymptotically flat region of the spacetime.
The frequency shift $g$ between the emission and detection 
of a photon is given by $g = E_\infty / E_{\rm plasma}$ with
$E_{\rm plasma} = -p_\mu u^\mu$ with $u^\mu$ being the 
plasma's four-velocity. 

The photon of Figure \ref{intensity} is launched with an energy 
at infinity of $1.7$ keV (blue segment). The first scattering 
decreases the energy at infinity to $0.7$ keV (green segment).
As the photon travels against the direction of the spinning space time of the black hole, the photon is swept up by the spacetime co-rotating with 
the black hole and appears to an outside observer to turns around.
The photon subsequently scatters a second and third time, increasing 
its energy at infinity first to $28.6$ keV and then to $28.8$ keV. 
Subsequently, the photons escapes the system as shown by the red segment. 
The main energy gain comes from the scattering at a 
radial coordinate of 1.28$\,r_{\rm g}$. Here, the gravitational 
radius of the black hole is given by $r_{\rm g}\,=\,G\,M\,/\,c^2$ with 
$G$ being the gravitational constant, $M$ the black hole mass, and $c$ the speed of light. 
For comparison, in the equatorial plane of the black hole, the event horizon (point of no return) is at 1.06$\,r_{\rm g}$ and the ergosphere (the region in which the extreme frame dragging of the black hole forces radiation and matter to co-rotate with the black hole) extends from the event horizon to 2$\,r_{\rm g}$.

In the remainder of this section we discuss the nature 
of the key scattering processes, and the power source of the new component. Collisional Penrose processes push one of the particles into 
trajectories with negative energy at infinity, while another one 
receives the energy difference \citep{bsw,2018GReGr..50...77S}. 
When the negative energy particles fall into the black hole, energy is extracted from the black hole.  The photons in the power-law 
component experience similar, but less extreme scattering processes. 
We can see that from the following calculation. 

A particle with mass $m$ at the radial Boyer-Lindquist coordinate $r$ 
orbits the black hole with spin $a$ (we assume $a>0$) with the orbital 
frequency in dimensional units ($M=1$, $r=r/r_{\rm g}$):
\begin{equation}
\Omega(r)=(r^{3/2}+a)^{-1}
\end{equation}
and with the four momentum
\begin{equation}
p^\alpha=p^0\,(1,0,0,\Omega)
\label{e:p}
\end{equation}
where $p^0$ follows from $p^2=-m\,c^2$.
We can use this result to analyze the key interaction of the example above,
occurring at $r\,=\,1.28\,r_{\rm g}$. 
Neglecting the thermal energy of the scattering electron 
(which our simulations do as well), we use $p^\alpha$ from Equation (\ref{e:p}) to infer the electron's energy at infinity from Equation (\ref{e:i}) to be 348\,keV.
This energy is lower then the 511\,keV 
rest mass energy of the electron as it is 
deeply bound in the black hole's gravitational potential. For the simulated trajectory of our example, we use the results from {\tt kerrC} to infer that the scattering increases the photon's energy at 
infinity by 26.9\,keV.  As the sum of the energies at infinity are conserved through the scattering process, the electron leaves the scattering in a trajectory with 
an energy at infinity of $E_{\infty}\,=$ 348\,keV$-$26.9\,keV\,$=$ 321.1\,keV. This is less than its energy before the scattering, but still very much positive. This example is representative for most of the 
scatterings responsible for the power-law component.
We conclude that the new emission component results from 
the kinematics of scatterings similar to collisional Penrose processes, 
but with the plasma electrons surviving the scatterings with
positive energies at infinity.
Given that the photosphere of the accretion disk is highly collisional, the scattering electrons will thus not plunge towards the black hole, but will equilibrate with the other particles in the accretion disk. 
As no negative energy particles are involved in the generation of the power-law component, we infer that the accretion disk - rather than the black hole spin - is the prime energy source of the new emission component. 
Note that Penrose processes are 
restricted to the ergosphere \citep[see][and references therein]{2010PhRvL.104b1101J, 2012PhysRevLett.109l1101B}, while
less extreme scattering processes can occur also outside of the ergosphere.

\begin{figure}[h!]
\centering
\vspace*{0.4cm}
\includegraphics[width=0.47\textwidth]{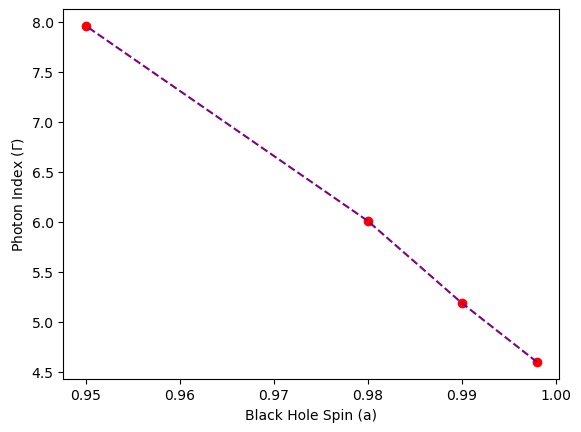}
\caption{Dependence of photon index on black hole spin for rapidly spinning ($0.95<a<0.998$) black holes. Black holes with lower spins have a ``steeper" index due to a smaller number of photons being boosted to higher energies.}\label{spectralindex}
\end{figure}

The power-law component becomes more prominent and ``harder'' (smaller $\Gamma$) for more rapidly spinning black holes ($a \ge 0.95$). This is due to the fact that the higher the spin, the closer the ISCO moves towards the black hole horizon, increasing the likelihood of photons emitted and reflected close to the black hole. Furthermore, the maximum energy gain per scattering increases for interactions occurring closer to the black hole, where both spacetime and the infalling matter move more rapidly and are under the influence of stronger frame dragging effects.
We show this impact of the black hole spin on the photon index in Figure \ref{spectralindex}. The spectrum is steeper (higher spectral index, e.g.\ $\Gamma\,\approx\,8$ for $a\,=\,0.95$) for less rapidly spinning black holes and harder (lower spectral index, e.g.\ $\Gamma\,\approx\,4.5$ for $a\,=\,0.998$) for extremely rapidly spinning black holes.
For $a\,\lesssim\,0.95$ the power law melts into the thermal energy spectrum and disappears.

\begin{figure*}
\centering
\includegraphics[width=0.5\textwidth]{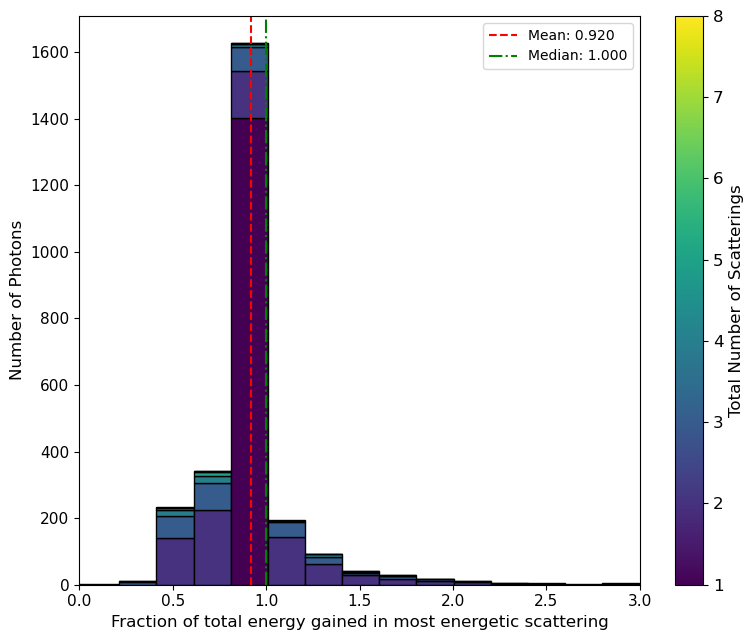}
\hfill
\includegraphics[width=0.49\textwidth]{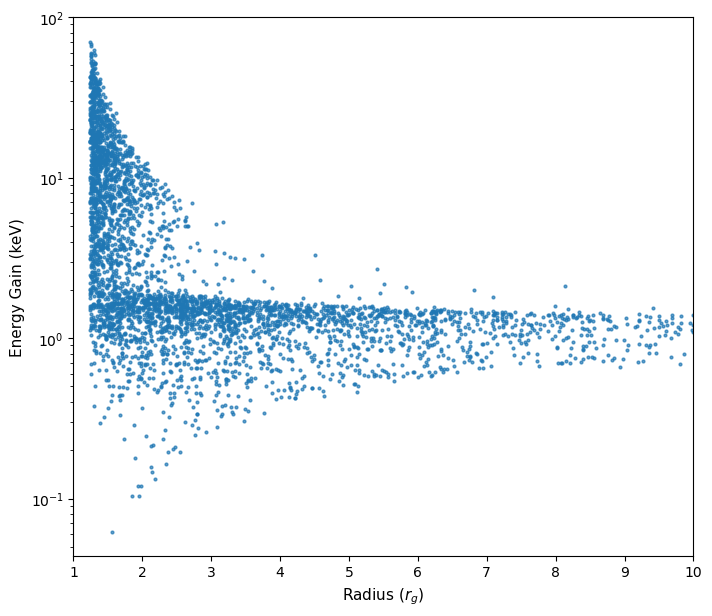}
\caption{Fraction of the final energy gained by power-law photons with $E_{\infty}>10$\,keV in the most energetic scattering (left), and the energy gain of photons in the power-law tail with $E_{\infty}>10$\,keV as a function of the radial coordinate of the scattering (right).  Only photons reaching an observer at an inclination of $33.6^{\circ}$ are shown. The ergosphere extends up to 2$r_{\rm g}$.}
\label{fig:tests}
\end{figure*}

We performed several checks to confirm that the result is genuine and does not stem from a numerical error.
Figure \ref{fig:tests} (left panel) shows the fraction of the final energy that the power-law photons with $E_{\infty}>10$\,keV gain in the most energetic scattering event.
We see that many photons scatter only once and receive most of their energy in a single scattering. Figure \ref{fig:tests} (right panel) shows the energy gain in keV as a function of the radial coordinate of the scattering.
Photons gain the most energy when they scatter within the ergosphere. The power law is composed of two photon populations: those coming from larger radii undergoing near head-on collisions with the material and acquiring large energy gains, and those that already arrive with energies $>$ 10 keV and are sent outward towards the observer.



Additionally, we verified the results with ray tracing simulations with enhanced numerical precision. We furthermore scrutinized individual photon trajectories, focusing on photons receiving large energy gains in an individual scattering. 
By raytracing them backward in time with improved precision and making use of the
well known quadrature integral equations 
for the geodesic motion, we verified that the photons receiving extreme energy gains indeed come from the accretion disk before receiving the energy gain. This excludes the possibility that numerical integration errors alone enable photons to get into trajectories that result in extreme energy gains. We also recalculated the tetrad transformations for processes leading to large energy gains with improved numerical accuracy and verified that the results remain unchanged. 
\section{Discussion}
\label{sec:4}
Our ray tracing studies show that the reflection of gravitationally lensed emission can give rise to a power-law component in the spectrum of the emission from a geometrically thin, optically thick accretion disk in the complete absence of a Comptonizing corona. As mentioned above, the power law 
largely results from individual scattering processes that leave the
scattering electrons in positive energy trajectories.
Our analysis shows that the most energetic processes tend to 
occur in the ergosphere, and thus probe the spacetime as well as the 
particle and photon kinematics in the ergosphere.

It should be noted that the appearance of the new emission component 
is not related to the electromagnetic energy extraction from a 
spinning black hole in the Blandford Znajek process \citep{1977MNRAS.179..433B,2011MNRAS.418L..94K}, 
nor to the driving of outflows in the Blandford-Payne process 
owing to centrifugally accelerated particles sliding along magnetic field lines anchored in the accretion disk \citep{1982MNRAS.199..883B}. 
We note that this paper, as all previous papers not based on first-principle
accretion disk simulations with radiative feedback, neglected the heating and cooling of the accretion disk from the scattering of the returning radiation. 
We will study the impact of the reflected returning radiation on the 
radial accretion disk temperature profile in a future paper.

The new component opens a new avenue for constraining black hole spins. 
Limits on the power-law component below the predictions can give upper limits on the spin. Spectropolarimetric observations of the new power-law component 
give new constraints on the black hole spin and inclination.
These constraints are complementary to other methods, i.e., the method of continuum fitting \citep{Zhang97} and inner accretion flow line fitting \citep{Reynolds2014}. Note that the composite multi-temperature 
disk emission and  relativistic power-law emission may explain some 
or all of the emission observed in the enigmatic soft power-law emission states, particularly for BHXRBs, in the Steep Power Law state (SPL) \citep{spl1,spl2} that exhibit power-law indices $\Gamma\,>\,2.4$.
Our simulations neglect additional scatterings in the plunging regions - the region between the ISCO and the black hole. We furthermore neglect any emission from the plunging region itself (e.g., \cite{2025MNRAS.541.3184H}, and references therein).  

The new emission component has several signatures that may allow to detect 
it unambiguously. (i) the composite multi-temperature blackbody plus reflected emission exhibits a distinct spectral shape of thermal emission rolling over into a power law without the transition from the fluxes from the thermal emission plunging into a harder power-law component. Detailed analysis of available data from hard X-ray observatories like \textit{NuSTAR} could help uncover these signatures, as its bandpass is well-suited to capture this smooth transition.
(ii) The power-law emission component originates from scattered photons, thus it
is strongly polarized. 
Figure \ref{spectrum} shows that the transition of the thermal to the reflected energy spectrum is accompanied by a 90$^{\circ}$ swing of the PA, and a strong increase of the PD to 8\% at an inclination of 33.6$^{\circ}$ - much higher than the PD expected for Comptonization alone. This leads to testable predictions in terms of the flux and polarization energy spectra.
The detection of the effect requires systems in which coronal 
Comptonization is too weak to mask the power-law component described here.
An accretion system that shows variations in the truncation radius over time would be particularly interesting to observe, as the power-law indices and PDs will follow a precise correlation. 

The observations require a combination of excellent spectral and polarization coverage in the energy range from a few keV to 20 keV or so. This could be accomplished by a joint observation with a {\it NuSTAR}-type mission with broadband spectral coverage, with a future broadband X-ray polarimeter \citep{PolSTAR,Soffitta} that extends the X-ray polarization beyond the 2-8\,keV energy of the {\it IXPE} satellite. From an observational standpoint, sources such as Cygnus X-1, LMC X-3, GX 339-4 and GRS 1915+105 are prime candidates due to their likely high spins, frequent soft states, and brightness. High-inclination systems are particularly favorable, as returning radiation contributes a larger fraction of the observed flux.


The emergence of the new power-law component relies on Kerr spacetime kinematics and Thomson scattering and is thus scale-invariant with respect to black hole mass. 
In principle, the same effect should produce power-law emission in Active Galactic Nuclei (AGNs) as well. 
However, detecting this power-law component in AGNs is difficult for several reasons.
The thermal thin disk emission from AGNs peaks in the optical and ultraviolet \citep{2012MNRAS.423..451L}, and the power-law would be observable in the extreme ultraviolet (EUV) which falls into the ``EUV gap’’ (caused by the absorption of EUV photons in the interstellar medium).
Furthermore, other, as yet not understood, emission components showing up in the ultra-soft X-ray band as the soft X-ray excess \citep{2023AN....34430105B}, will likely mask the power-law emission.  

The unambiguous detection of the emission from the ergospheres of BHXRBs is an exciting new chapter in our exploration of black holes (see also \cite{2025MNRAS.544.2880M} and references therein) and is another channel for testing predictions of the theory of General Relativity in the strong gravity regime.

\begin{acknowledgments}
The authors thank Alex Chen, Yajie Yuan, John Mehlhaff, Aaron Barrios, and Robert Heyck for interesting discussions 
of the Blandford Znajek effect, and Chris Done and Michael Nowak 
for emphasizing the importance of the heating of the disk 
by the Compton scattering of returning emission.  
The authors acknowledge salary support through the grants 
80NSSC24K0205, 80NSSC24K1178,80NSSC24K1749, and 80NSSC24K1819, 
as well as from the McDonnell Center for the Space Sciences.
\end{acknowledgments}

\begin{contribution}

All authors made significant contributions to the development of the paper and are considered equal. 
SVM performed the data analysis, wrote the initial draft, and performed the tests of the nature of the effect. KH and HK developed the {\tt{kerrC}} code, found the first evidence for the effect, and contributed to the 
interpretation of the results and with the writing of the paper. 


\end{contribution}

\section*{Code Availability}

The {\tt{kerrC}} code and the analysis code used for this paper are published in the linked Zenodo repository: \url{https://doi.org/10.5281/zenodo.19711393}.\\

\bibliography{ref}{}
\bibliographystyle{aasjournalv7}



\end{document}